%% file: cikm2016-latent-entities.tex
\pdfoutput=1

\documentclass{sig-alternate-05-2015}

\CopyrightYear{2016}
\setcopyright{acmlicensed}
\conferenceinfo{CIKM'16 ,}{October 24 - 28, 2016, Indianapolis, IN, USA}
\isbn{978-1-4503-4073-1/16/10}\acmPrice{\$15.00}
\doi{http://dx.doi.org/10.1145/2983323.2983702}

\usepackage[
  pass,
]{geometry}

\usepackage{amsmath}
\usepackage{booktabs}
\usepackage{color}
\usepackage{enumitem}
\usepackage{flushend}
\usepackage{footnote}
\usepackage{hyperref}
\usepackage{grffile}
\usepackage{mathtools}
\DeclarePairedDelimiter{\ceil}{\lceil}{\rceil}
\usepackage{multirow}
\usepackage[all]{nowidow}
\usepackage[np, autolanguage]{numprint}
\usepackage{setspace}
\usepackage{subfig}
\usepackage{times}
\usepackage{tikz}
\usetikzlibrary{decorations.markings, calc, fadings,
decorations.pathreplacing, patterns, decorations.pathmorphing, positioning, shapes.misc}
\tikzset{
	cross/.style = {cross out, draw,
					minimum size=2*(#1-\pgflinewidth),
         			inner sep=0pt, outer sep=0pt},
    cross/.default={1pt}
}
\usepackage{tikz-3dplot}
\usepackage{todonotes}
\usepackage[square,comma,numbers,sort&compress,sectionbib]{natbib}
\usepackage{paralist}
\usepackage[labelfont=bf,textfont=bf,skip=0pt]{caption}
\usepackage{etoolbox}
\usepackage{array}
\newcolumntype{L}[1]{>{\raggedright\let\newline\\\arraybackslash\hspace{0pt}}m{#1}}
\newcolumntype{C}[1]{>{\centering\let\newline\\\arraybackslash\hspace{0pt}}m{#1}}
\newcolumntype{R}[1]{>{\raggedleft\let\newline\\\arraybackslash\hspace{0pt}}m{#1}}

\hypersetup{pdfinfo={
Title={Learning Latent Vector Spaces for Product Search},
Author={Christophe Van Gysel, Maarten de Rijke, Evangelos Kanoulas}
}}

\makeatletter
\renewcommand{\paragraph}{%
    \@startsection{paragraph}{4}{\z@}{0\p@ \@plus \p@}
    {-5\p@}{\subsecfnt}%
}
\makeatother

\def \paperImplementationUrl {\url{https://github.com/cvangysel/SERT}}

\def \ModelName{LSE}
\def \FullModelName{Latent Semantic Entities}

\def \QueryLikelihoodLM{Query-likelihood Language Model}
\def \QLM{QLM}

\def \LDA{LDA}
\def \LSI{LSI}
\def \WordToVec{word2vec}

\def \RankSVM{RankSVM}

\def \ResearchQuestionOne {How do the parameters of \ModelName{} influence its efficacy?}
\def \ResearchQuestionTwo {How does \ModelName{} compare to latent vector models based on \LDA{}, \LSI{} and \WordToVec{}?}
\def \ResearchQuestionThree {How does \ModelName{} compare to a smoothed language model that applies lexical term matching?}
\def \ResearchQuestionFour {What is the benefit of incorporating \ModelName{} as a feature in a learning-to-rank setting?}

\def \ContributionOne {A latent vector model, \ModelName{}, that jointly learns the representations of words, entities and the relationship between the former, together with an open-source implementation.\footnote{\paperImplementationUrl}}
\def \ContributionTwo {A study of the influence of \ModelName{}'s parameters and how these influence its ability to discriminate between entities.}
\def \ContributionThree {An in-depth comparative analysis of the entity retrieval effectiveness of latent vector models.}
\def \ContributionFour {Insights in how \ModelName{} can improve retrieval performance in entity-oriented search engines.}
\def \ContributionFive {An analysis of the differences in performance between latent vector models by examining entity representations and mappings from queries to entity space.}

\newcommand{\MultiRowCell}[1]{\begin{tabular}{r}#1\end{tabular}}

\newcommand{\norm}[1]{\left\lVert#1\right\rVert}

\makeatletter
\newcolumntype{"}{@{\hskip\tabcolsep\vrule width 1pt\hskip\tabcolsep}}
\makeatother

\allowdisplaybreaks

\begin{document}

\title{Learning Latent Vector Spaces for Product Search}

\numberofauthors{1}
\author{
\alignauthor
\mbox{}\hfill
\begin{tabular}{c}
    Christophe Van Gysel\\
   \email{cvangysel@uva.nl}
\end{tabular}
\hfill
\begin{tabular}{c}
    Maarten de Rijke\\
   \email{derijke@uva.nl}
\end{tabular}
\hfill
\begin{tabular}{c}
    Evangelos Kanoulas\\
   \email{e.kanoulas@uva.nl}
\end{tabular}
\hfill\mbox{}\\[1.1ex]
   \affaddr{University of Amsterdam, Amsterdam, The Netherlands}
}

\maketitle
\begin{abstract}
We introduce a novel latent vector space model that jointly learns the latent representations of words, e-commerce products and a mapping between the two without the need for explicit annotations. The power of the model lies in its ability to directly model the discriminative relation between products and a particular word. We compare our method to existing latent vector space models (LSI, LDA and word2vec) and evaluate it as a feature in a learning to rank setting. Our latent vector space model achieves its enhanced performance as it learns better product representations. Furthermore, the mapping from words to products and the representations of words benefit directly from the errors propagated back from the product representations during parameter estimation. We provide an in-depth analysis of the performance of our model and analyze the structure of the learned representations.
\end{abstract}

%
%

\ccsdesc[500]{Information systems~Content analysis and feature selection}

%
%

%
%

\subsection*{Keywords}{Entity retrieval; Latent space models; Representation learning}

\input{introduction}
\input{related_work}
\input{methodology}
\input{experiments}
\input{discussion}
\input{conclusions}

\vspace*{-1.0em}
\renewcommand{\bibsection}{
\section*{REFERENCES}%
\vspace*{-0.5\baselineskip}
}
\setlength{\bibsep}{0pt}
\bibliographystyle{abbrvnatnourl}
{
 \raggedright\small
 \bibliography{cikm2016-latent-entities}
}
{
\small%
\vspace*{-1.0em}
\input{appendix}
}

\end{document}

%% file: introduction.tex

\section{Introduction}

Retail through online channels has become an integral part of consumers' lives \citep{PricewaterhouseCoopers}. In addition to using these online platforms that generate hundreds of billions of dollars in revenue \citep{Forrester}, consumers increasingly participate in multichannel shopping where they research items online before purchasing them in brick-and-mortar stores. Search engines are essential for consumers to be able to make sense of these large collections of products available online \citep{Jansen2006}. In the case of directed searching (in contrast to exploratory browsing), users formulate queries using characteristics of the product they are interested in (e.g., terms that describe the product's category) \citep{Rowley2000}. However, it is widely known that there exists a mismatch between queries and product representations where both use different terms to describe the same concepts \citep{Li2014}. Thus, there is an urgent need for better semantic matching methods.
\vfill
\pagebreak

Product search is a particular example of the more general entity finding task that is increasingly being studied. Other entity finding tasks considered recently include searching for people~\citep{Balog2012}, books~\citep{Gade2015} and groups~\citep{Liang2016}. Products are retrievable entities where every product is associated with a description and one or more user reviews. Therefore, we use the terms ``product'' and ``entity'' interchangeably in this paper. However, there are two important differences between product search and the entity finding task as defined by \citet{DeVries2007}. First, in entity finding one retrieves entities of a particular type from large broad coverage multi-domain knowledge bases such as Wikipedia \citep{DeVries2007,Balog2012dbpedia}. In contrast, product search engines operate within a single domain which can greatly vary in size. Second, user queries in product search consist of free-form text \citep{Rowley2000}, as opposed to the semi-structured queries with additional type or relational constraints being used in entity finding~\citep{DeVries2007,Balog2011}.

In this paper we tackle the problem of discriminating between products based on the language (i.e., descriptions and reviews) they are associated with. Existing methods that are aimed at discriminating between entities based on textual data learn word representations using a language modeling objective or heuristically construct entity representations \citep{VanGysel2016,Demartini2009}. Our approach directly learns two things: a unidirectional mapping between words and entities, as well as distributed representations of both words and entities. It does so in an unsupervised and automatic manner such that words that are strongly evidential for particular products are projected nearby those products.
While engineering of representations is important in information retrieval \citep{Balog2007,Demartini2009,Bordes2011,Cai2015,Zhao2015,Graus2016}, unsupervised joint representation learning of words and entities has not received much attention.
We fill this gap.
Our focus on learning representations for an end-to-end task such as \emph{product search} is in contrast to the large volume of recent literature on word representation learning \citep{Turian2010} that has a strong focus on upstream components such as distributional semantics \citep{Mikolov2013,Pennington2014}, parsing \citep{Turian2010,Collobert2011} and information extraction \citep{Turian2010,Collobert2011}.
In addition, our focus on \emph{unsupervised} representation learning is in contrast to recent entity representation learning methods \citep{Bordes2011,Zhao2015} that heavily depend on precomputed entity relationships and cannot be applied in their absence.

In recent years, significant progress has been made concerning semantic representations of entities. We point out three key insights on which we build:
\begin{inparaenum}[(1)]
	\item Distributed representations \citep{Hinton1986} learned by discriminative neural networks reduce the curse of dimensionality and improve generalization. Latent features encapsulated by the model are shared by different concepts and, consequently, knowledge about one concept influences knowledge about others.
	\item Discriminative approaches outperform generative models if enough training data is available \citep{Ng2002,Baroni2014} as discriminative models solve the classification problem directly instead of solving a more general problem first \citep{Vapnik1998}.
	\item Recently proposed unsupervised neural retrieval models \citep{VanGysel2016} do not scale as they model a distribution over all retrievable entities; the approach is infeasible during training if the collection of retrievable entities is large.
\end{inparaenum}%

Building on these insights, we introduce \FullModelName{} (\ModelName{}), a method that learns separate representations of words and retrievable objects jointly for the case where mostly unstructured documents are associated with the objects (i.e., descriptions and user reviews for products) and without relying on predefined relationships between objects (e.g., knowledge graphs). \ModelName{} learns to discriminate between entities for a given word sequence by mapping the sequence into the entity representation space. Contrary to heuristically constructed entity representations \citep{Demartini2009}, \ModelName{} learns the relationship between words and entities directly using gradient descent. Unlike \citep{VanGysel2016}, we avoid computing the full probability distribution over entities; we do so by using noise-contrastive estimation.

Our research questions are as follows:
\begin{inparaenum}[(1)]
	\item \ResearchQuestionOne{}
	\item \ResearchQuestionTwo{}
	\item \ResearchQuestionThree{}
	\item \ResearchQuestionFour{}
\end{inparaenum}

We contribute:
\begin{inparaenum}[(1)]
	\item \ContributionOne{}
	\item \ContributionTwo{}
	\item \ContributionThree{}
	\item \ContributionFour{}
	\item \ContributionFive{}
\end{inparaenum}

%% file: related_work.tex

\vspace*{-0.25em}
\section{Related work}
\label{section:related_work}

\vspace*{-0.25em}
\subsection{Product retrieval}

Product search engines are an important source of traffic in the e-commerce market \citep{Jansen2006}. Specialized solutions are needed to maximize the utilization of these platforms. \citet{Nurmi2008} note a discrepancy between buyers' shopping lists and how retail stores maintain information. They introduce a grocery retrieval system that retrieves products using shopping lists written in natural language. Product resolution \citep{Balog2011-2} is an important task for e-commerce aggregation platforms, such as verticals of major web search engines and price comparison websites. \citet{Duan2013} propose a probabilistic mixture model for the attribute-level analysis of product search logs. They focus on structured aspects of product entities, while in this work we learn representations from unstructured documents. \citet{Duan2013-2} extend the language modeling approach to product databases by incorporating the ability to condition on specification (e.g., lightweight products only). They note that while languages such as SQL can be used effectively to query these databases, their use is difficult for non-experienced end users. \citet{Duan2015} study the problem of learning query intent representation for structured product entities. They emphasize that existing methods focus only on the query space and overlook critical information from the entity space and the connection in between. 

We agree that modeling the connection between query words and entities and propagating information from the entity representations back to words is essential. In contrast to their work, we consider the problem of learning representations for entities based on their associations with unstructured documents.

\subsection{Latent semantic information retrieval}

The mismatch between queries and documents is a critical challenge in search \citep{Li2014}. Latent Semantic Models (LSMs) enable retrieval based on conceptual content, instead of exact word matches. LSMs have become popular through the introduction of Latent Semantic Indexing (LSI) \citep{Deerwester1990}, followed by probabilistic LSI (pLSI) \citep{Hofmann1999}. \citet{Salakhutdinov2009} use a deep auto-encoder for the unsupervised learning of latent semantic document bit patterns. Deep Structured Semantic Models \citep{Huang2013,Shen2014} employ click data to predict a document's relevance to a query. Methods based on neural networks have also been used for machine-learned ranking \citep{Burges2005,Deng2013,Liu2011}. \citet{VanGysel2016} introduce an LSM for entity retrieval, with an emphasis on expert finding; they remark that training the parameters of their model becomes infeasible when the number of entities increases. In this work we mitigate this problem by considering only a random sample of entities as negative examples during training. This allows us to efficiently estimate model parameters in large product retrieval collections, which is not possibly using the approach of \citep{VanGysel2016} due to its requirement to compute a normalization constant over all entities.
\subsection{Representation learning}

Recently, there has been a growing interest in neural probabilistic language models (LMs) for the modeling of word sequences \citep{Bengio2003,Mnih2007,Mnih2009}. Distributed representations \citep{Hinton1986} of words learned by neural LMs, also known as word embeddings, incorporate syntactic and semantic information \citep{Mnih2013,Mikolov2013,Pennington2014} as a side-effect of their ability to reduce the dimensionality. Feed-forward \citep{Collobert2011} and recurrent \citep{Mikolov2013} neural networks perform well in various NLP tasks. Very recently, there has been an increased interest in multimodal neural language models \citep{Kiros2014}, which are used for the task of automated image captioning, amongst others.
Learning representations of entities is not new. \citet{Bordes2011} leverage structured relations captured in Knowledge Bases (KB) for entity representation learning and evaluate their representations on the link prediction task. Our approach has a strong focus on modeling the language of all entities collaboratively, without the need for explicit entity relations during training. \citet{Zhao2015} employ matrix factorization methods to construct low-dimensional continuous representations of entities, categories and words for determining similarity of Wikipedia entities. They employ a word pair similarity evaluation set and only evaluate on pairs referring to Wikipedia entities; they learn a single semantic space for widely-differing concepts (entities, categories and words) of different cardinalities and make extensive use of an underlying Knowledge Graph (KG) to initialize their parameters. In contrast, we model representations of words and entities jointly in separate spaces, in addition to a mapping from word to entity representations, in an unsupervised manner.

We tackle the task of learning latent continuous vector representations for e-commerce products for the purpose of product search. The focus of this work lies in the language modeling and representation learning challenge. We learn distributed representations \citep{Hinton1986} of words and entities and a mapping between the two. At retrieval time, we rank entities according to the similarity of their latent representations to the projected representation of a query. Our model \ModelName{} is compared against existing entity-oriented latent vector representations that have been created using \LSI{}, \LDA{} and \WordToVec{}. We provide an analysis of model parameters and give insight in the quality of the joint representation space.

%% file: methodology.tex

\vfill

\section{Latent vector spaces for\\entity retrieval}
\label{section:methodology}

\newcommand{\Prob}[1]{P(#1)}
\newcommand{\CondProb}[2]{\Prob{#1 \mid #2}}

\newcommand{\UnnormedProb}[1]{\tilde{P}(#1)}
\newcommand{\UnnormedCondProb}[2]{\UnnormedProb{#1 \mid #2}}

\newcommand{\SigmoidFn}[1]{\sigma{}(#1)}
\newcommand{\ExpFn}[1]{e^{#1}}

\newcommand{\DotProduct}[2]{#1 \cdot #2}
\newcommand{\Transpose}[1]{#1^\intercal}

\newcommand{\Length}[1]{|#1|}

\newcommand{\Entities}{X}
\newcommand{\Entity}{\MakeLowercase{\Entities{}}}

\newcommand{\NumEntities}{\Length{\Entities{}}}

\newcommand{\Vocabulary}{V}
\newcommand{\WordEmbedding}{\MakeLowercase{\Vocabulary}}

\newcommand{\VocabularySize}{\Length{\Vocabulary{}}}

\newcommand{\EntitySpace}{E}
\newcommand{\EntityEmbedding}{\MakeLowercase{\EntitySpace}}

\newcommand{\String}{s}
\newcommand{\Word}{w}

\newcommand{\SimilarityFn}{S_c}

\newcommand{\HashFn}{f}

\newcommand{\WordEmbeddingSize}{e_\Vocabulary{}}
\newcommand{\EntityEmbeddingSize}{e_\EntitySpace{}}

\newcommand{\VocabularyEmbeddingMatrix}{W_\WordEmbedding{}}
\newcommand{\MappingMatrix}{W}
\newcommand{\MappingBias}{b}
\newcommand{\EntityEmbeddingMatrix}{W_\EntityEmbedding{}}

\newcommand{\CategoricalTopic}{c}
\newcommand{\Query}{q}
\newcommand{\Term}{t}

\newcommand{\Beacon}{\tilde{\EntityEmbedding{}}}

\newcommand{\Rank}{\text{rank}}

\newcommand{\Documents}{D}
\newcommand{\Document}{\MakeLowercase{\Documents{}}}
\newcommand{\DocumentEmbedding}[1][]{\HashFn{}({\Document{}#1})}

\newcommand{\AssociatedDocuments}[1]{\Documents{}_{#1}}

\newcommand{\WindowSize}{n}

\newcommand{\NumNegativeExamples}{z}

\newcommand{\BatchSize}{m}

\newcommand{\LossFnIdentifier}{L}
\newcommand{\LossFnArguments}[1][]{(\VocabularyEmbeddingMatrix{}#1, \EntityEmbeddingMatrix{}#1, \MappingMatrix{}#1, \MappingBias{}#1)}
\newcommand{\LossFn}[1][]{\LossFnIdentifier{}\LossFnArguments[#1]{}}

\newcommand{\LearningRate}{\boldsymbol{\alpha}}

\tdplotsetmaincoords{0}{0}
\newcommand{\TwoDAxes}[1][]{
\begin{tikzpicture}
    [scale=3,
     tdplot_main_coords,
     axis/.style={->,black,thick}]

\coordinate (O) at (0,0);

\draw (0,0,0) -- (1,0,0);
\draw (0,0,0) -- (0,1,0);
\draw (1,0,0) -- (1,1,0);
\draw (0,1,0) -- (1,1,0);
#1
\end{tikzpicture}
}

\begin{figure}
	\centering
	\begin{tikzpicture}[>=latex]%
		\node[cm={1,0,cos(20),sin(20),(0,0)}] (MetricSpace) {
			\TwoDAxes[
				\foreach \i in {1,...,9} {
					\draw[red] (0.5,0.5,0) circle (0.5*0.05cm^\i);
				}
				\foreach \i in {(0.48, 0.40), (0.75,0.75), (0.1,0.9)} {
					\draw \i node[cross=5pt,rotate=0,line width=0.75mm] {};
				}
			]{}
		};
		\draw[->] (-0.877582562,0.4794255386) to [bend left=50] (MetricSpace.center);
	\end{tikzpicture}

        \medskip
	\caption{Illustrative example of how entities are ranked in vector space models w.r.t. a projected query. Query $\Query{}$ is projected into entity space $\EntitySpace{}$ using mapping $\HashFn{}$ (black arrow) and entities (black crosses) are ranked according to their similarity in decreasing order.\label{fig:inference}}
\end{figure}
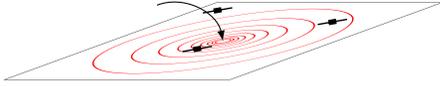

\begin{figure*}
	\newcommand{\PrettyVector}[1]{
		\left[
	    \begin{matrix}
	     	#1
	    \end{matrix}
		\right]
	}
	\newcommand{\PrettyOneHotVector}{\PrettyVector{\\ 0 \\ \vdots{} \\ 1 \\ \vdots{} \\ 0 \\ \\}}
	\newcommand{\PrettyFilledVector}[1]{\PrettyVector{\\ \\ \\ #1 \\ \\ \\ \\}}
	\newcommand{\Sized}[2]{#1 \left\{ #2 \right.}
	\resizebox{\textwidth}{!}{
	\begin{tikzpicture}[>=latex]%
		\node[draw=white,line width=0] (Vocabulary) {$\Word \in \Vocabulary{}$};
		\node[draw=white,line width=0] (VocabularyEmbedding) [right=2cm of Vocabulary] {$\Sized{\WordEmbeddingSize{}}{\PrettyFilledVector{\WordEmbedding{}}}$};
		\node[draw=white,line width=0] (WordBeacon) [right=2.0cm of VocabularyEmbedding] {$\Sized{\EntityEmbeddingSize{}}{\PrettyFilledVector{\Beacon{}}}$};

		\node[cm={1,0,cos(20),sin(20),(0,0)}] (MetricSpace) [right=1.5cm of WordBeacon] {\TwoDAxes{}};

		\node[draw=white,line width=0] (EntityEmbedding) [right=1.5cm of MetricSpace] {$\Sized{\EntityEmbeddingSize{}}{\PrettyFilledVector{\EntityEmbedding{}}}$};
		\node[draw=white,line width=0] (Entity) [right=2cm of EntityEmbedding] {$\Entity \in \EntitySpace{}$};

		\draw[->] (Vocabulary) -> (VocabularyEmbedding) node[above,midway,text width=2cm,align=center] {Look-up embedding in\\$\VocabularyEmbeddingMatrix{}$};
		\draw[->] (VocabularyEmbedding) -> (WordBeacon) node[above,midway,text width=2cm, align=center] {Transform with\\\scriptsize{$\tanh{({\MappingMatrix{} \cdot \WordEmbedding{} + \MappingBias{})}}$}};
		\draw[->] (Entity) -> (EntityEmbedding) node[above,midway,text width=2cm,align=center] {Look-up embedding in\\$\EntityEmbeddingMatrix{}$};

		\draw[->] (EntityEmbedding) to [bend right=50] ($(MetricSpace.center)!0.50!(MetricSpace.north east)$);
		\draw[->] (WordBeacon) to [bend left=50] ($(MetricSpace.center)!0.50!(MetricSpace.south west)$);

		\begin{scope}[transparency group, opacity=0.75]
		\draw[<->,cyan] ($(MetricSpace.center)!0.50!(MetricSpace.south west)$) -- node[below=0,midway,black] {$\SimilarityFn{}(\Beacon{}, \EntityEmbedding{})$} ($(MetricSpace.center)!0.50!(MetricSpace.north east)$);
		\end{scope}

		\draw[decorate,decoration={brace,amplitude=10pt}]
([yshift=1.5cm]Vocabulary.west) -- ([yshift=1.5cm]WordBeacon.east) node [black,midway,yshift=0.75cm] {$\HashFn{}(\Word{})$};
	\end{tikzpicture}}

	\caption{Schematic representation of the \FullModelName{} model for a single word $\Word{}$. Word embeddings $\VocabularyEmbeddingMatrix{}$ ($\WordEmbeddingSize{}$-dim. for $\VocabularySize{}$ words), entity embeddings $\EntityEmbeddingMatrix{}$ ($\EntityEmbeddingSize{}$-dim. for $\NumEntities{}$ entities) and the mapping from words to entities ($\EntityEmbeddingSize{}$-by-$\WordEmbeddingSize{}$ matrix $\MappingMatrix{}$, $\EntityEmbeddingSize{}$-dim. vector $\MappingBias{}$) are learned using gradient descent.\label{fig:schema}}
\end{figure*}

We first introduce a generalized formalism and notation for entity-oriented latent vector space models. After that, in \S\ref{section:methodology:model}, we introduce \FullModelName{}, a latent vector space model that jointly learns representations of words, entities and a mapping between the two directly, based on the idea that entities are characterized by the words they are associated with and vice versa. Product representations are constructed based on the n-grams the products are likely to generate based on their description and reviews, while word representations are based on the entities they are associated with and the context they appear in. We model the relation between word and product representations explicitly so that we can predict the product representation for a previously unseen word sequence.

\subsection{Background}
\label{section:methodology:background}

\newcommand{\ProjectedQuery}[1][]{\HashFn{}(\Query{}#1)}

We focus on a product retrieval setting in which a user wants to retrieve the most relevant products on an e-commerce platform. As in typical information retrieval scenarios, the user encodes their information need as a query $\Query{}$ and submits it to a search engine. Product search queries describe characteristics of the product the user is searching for, such as a set of terms that describe the product's category \citep{Rowley2000}.

Below, $\Entities{}$ denotes the set of entities that we consider. For every $\Entity{}_i \in \Entities{}$ we assume to have a set of associated documents $\AssociatedDocuments{\Entity{}_i}$. The exact relation between the entity and its documents depends on the problem setting. 
In this paper, entities are products \citep{Nurmi2008,Duan2015} and documents associated with these products are descriptions and product reviews.

Latent vector space models rely on a function $\HashFn{}: \Vocabulary{}^+ \to \EntitySpace{}$ that maps a sequence of words (e.g., a query $\Query{}$ during retrieval) from a vocabulary $\Vocabulary{}$ to a $\EntityEmbeddingSize{}$-dimensional continuous entity vector space $\EntitySpace{} \subset \mathbb{R}^{\EntityEmbeddingSize{}}$. Every entity $\Entity{}_i \in \Entities{}$ has a corresponding vector representation $\EntityEmbedding{}_i \in \EntitySpace{}$. Let $\SimilarityFn{}: \EntitySpace{} \times \EntitySpace{} \to \mathbb{R}$ denote the cosine similarity between vectors in $\EntitySpace{}$. For a given query $\Query{}$, entities $\Entity{}_i$ are ranked in decreasing order of the cosine similarity between $\EntityEmbedding{}_i$ and the query projected into the space of entities, $\ProjectedQuery{}$. Fig.~\ref{fig:inference} illustrates how entities are ranked according to a projected query. For \LSI{}, $\HashFn{}$ is defined as the multiplication of the term-frequency vector representation of $\Query{}$ with the rank-reduced term-concept matrix and the inverse of the rank-reduced singular value matrix \citep{Deerwester1990}. In the case of \LDA{}, $\HashFn{}$ becomes the distribution over topics conditioned on $\Query{}$ \citep{Blei2013}. This distribution is computed as the sum of the topic distributions conditioned on the individual words of $\Query{}$. In this paper, the embedding $\HashFn{}$ is learned; see \S\ref{section:methodology:estimation} below.

Traditional vector space models operate on documents instead of entities. \citet{Demartini2009} extend document-oriented vector spaces to entities by representing an entity as a weighted sum of the representations of their associated documents:
\begin{equation}
\label{eq:demartini}
\EntityEmbedding{}_i = \sum_{\Document{}_j \in \AssociatedDocuments{\Entity{}_i}} r_{i,j} \DocumentEmbedding[_j]{}
\end{equation}
where $\DocumentEmbedding[_j]{}$ is the vector representation of $\Document{}_j$ and $r_{i,j}$ denotes the relationship weight between document $\Document{}_j$ and entity $\Entity{}_i$. In this work we put $r_{i,j} = 1$ whenever $\Document{}_j \in \AssociatedDocuments{\Entity{}_i}$ for a particular $\Entity{}_i \in \Entities{}$ and $r_{i,j}=0$ otherwise, as determining the relationship weight between entities and documents is a task in itself.

\subsection{Latent semantic entities}
\label{section:methodology:model}

While Eq.~\ref{eq:demartini} adapts document-oriented vector space models to entities, in this work we define $\HashFn{}$ by explicitly learning (\S\ref{section:methodology:estimation}) the mapping between word and entity representations and the representations themselves:
\newcommand{\OneHotVector}{\delta{}_i}
\newcommand{\NormalizedBoWVector}{\frac{1}{\Length{\String{}}} \sum_{\Word{}_i \in \String{}} \OneHotVector{}}
\newcommand{\AvgWordEmbedding}{\VocabularyEmbeddingMatrix{} \cdot \NormalizedBoWVector{}}
\newcommand{\LinearCombination}{\MappingMatrix{} \cdot (\AvgWordEmbedding{}) + \MappingBias{}}
\begin{equation}
\label{eq:hash_fn}
\HashFn{}(\String{}) = \tanh\left(\LinearCombination{}\right)
\end{equation}
for a string $\String{}$ of constituent words $\Word{}_1, \ldots, \Word{}_{|\String{}|}$ (an n-gram extracted from a document or a user-issued query), where $\VocabularyEmbeddingMatrix{}$ is the $\WordEmbeddingSize{} \times \VocabularySize{}$ projection matrix that maps the averaged one-hot representations (i.e., a $|V|$-dimensional vector with element $i$ turned on and zero elsewhere) of word $\Word{}_i$, $\OneHotVector{}$, to its $\WordEmbeddingSize{}$-dimensional distributed representation. This is equivalent to taking the embeddings of the words in $\String{}$ and averaging them. In addition, $\MappingBias{}$ is a $\EntityEmbeddingSize{}$-dimensional bias vector, $\MappingMatrix{}$ is the $\EntityEmbeddingSize{} \times \WordEmbeddingSize{}$ matrix that maps averaged word embeddings to their corresponding position in entity space $\EntitySpace{}$ and $\tanh$ is the element-wise smooth hyperbolic tangent with range $(-1, 1)$. This transformation allows word embeddings and entity embeddings to be of a different dimensionality.

In other words, for a given string of words we take the representation of this string to be the average of the representations of the words it contains \citep{Mikolov2013,Le2014}. This averaged word representation is then transformed using a linear map ($\MappingMatrix{}$) and afterwards translated using $\MappingBias{}$. We then apply the hyperbolic tangent as non-linearity such that every component lies between $-1$ and $1$. First of all, this regularizes the domain of the space and avoids numerical instability issues that occur when the magnitude of the vector components becomes too large. Secondly, by making the function non-linear we are able to model non-linear class boundaries in the optimization objective that we introduce in the next section. We use $\EntityEmbeddingMatrix{}$ to denote the $|\Entities{}| \times \EntityEmbeddingSize{}$ matrix that holds the entity representations. Row $i$ of $\EntityEmbeddingMatrix{}$ corresponds to the vector representation, $\EntityEmbedding{}_i$, of entity $\Entity{}_i$. Fig.~\ref{fig:schema} depicts a schematic overview of the proposed model. The parameters $\VocabularyEmbeddingMatrix{}$, $\MappingMatrix{}$, $\MappingBias{}$ and $\EntityEmbeddingMatrix{}$ will be learned automatically using function approximation methods as explained below.

The model proposed in this section shares similarities with previous work on word embeddings and unsupervised neural retrieval models \citep{Mikolov2013,VanGysel2016}. However, its novelty lies in its ability to scale to large collections of entities and its underlying assumption that words and entities are embedded in spaces of different dimensionality:
\begin{inparaenum}[(1)]
	\item The model of \citep{Mikolov2013} has no notion of entity retrieval as it estimates a language model for the whole corpus.
	\item Similar to \citep{Mikolov2013}, Eq.~\ref{eq:hash_fn} aggregates words $\Word{}_i \in \String{}$ to create a single phrase representation of $\String{}$. However, in \citep{VanGysel2016}, a distribution $\CondProb{\Entities{}}{\Word{}_i}$ is computed for every $\Word{}_i$ independently and aggregation occurs using the factor product. This is infeasible during model training when the collection of retrievable objects becomes too large, as is the case for product search. In the next section (\S\ref{section:methodology:estimation}) we solve this problem by sampling.
	\item In both \citep{Mikolov2013,VanGysel2016} two sets of representations of the same dimensionality are learned for different types of objects with potentially different latent structures (e.g., words, word contexts and experts). As mentioned earlier, Eq.~\ref{eq:hash_fn} alleviates this problem by transforming one latent space to the other.
\end{inparaenum}

\subsection{Parameter estimation}
\label{section:methodology:estimation}

\newcommand{\NGram}[2][]{#1{\Word{}_{#21}}, \ldots, #1{\Word{}_{#2\WindowSize{}}}}
\newcommand{\NGramProjection}{\HashFn{}({\NGram{}})}

\newcommand{\IndexedNGram}{\NGram[]{j,}}
\newcommand{\IndexedNGramProjection}{\HashFn{}({\IndexedNGram{}})}

\newcommand{\TargetEntity}{\Entity{}_i}
\newcommand{\TargetEntityRepr}{\EntityEmbedding{}_i}

For a particular document $\Document \in \AssociatedDocuments{\Entity{}_i}$ associated with entity $\TargetEntity{}$, we generate n-grams $\IndexedNGram{}$ where $\WindowSize{}$ (window size) remains fixed during training. For every n-gram $\IndexedNGram{}$, we compute its projected representation $\IndexedNGramProjection{}$ in $\EntitySpace{}$ using $\HashFn{}$ (Eq.~\ref{eq:hash_fn}). The objective, then, is to directly maximize the similarity between the vector representation of the entity $\EntityEmbedding{}_i$ and the projected n-gram $f(w_{j,1}$, \ldots, $w_{j,n})$ with respect to $\SimilarityFn{}$ (\S\ref{section:methodology:background}), while minimizing the similarity between $\IndexedNGramProjection{}$ and the representations of non-associated entities. This allows the model to learn relations between neighboring words in addition to the associated entity and every word.

However, considering the full set of entities for the purpose of discriminative training can be costly when the number of entities $\NumEntities{}$ is large. Therefore, we apply a variant of Noise-Contrastive Estimation (NCE) \citep{Gutmann2010,Mnih2012,Mnih2013,Mikolov2013-2} where we sample negative instances from a noise distribution with replacement. We use the uniform distribution over entities as noise distribution. Define
\newcommand{\SimilarityRandomVariable}{\mathcal{S}}
\newcommand{\ProbSimilarEntity}[1]{\CondProb{\SimilarityRandomVariable{}}{#1, \IndexedNGramProjection{}}}
\newcommand{\ProbMassEntity}[1]{\SigmoidFn{\DotProduct{#1}{\IndexedNGramProjection{}}}}
\begin{equation}
\ProbSimilarEntity{\TargetEntityRepr{}} = \ProbMassEntity{\TargetEntityRepr{}}
\end{equation}
as the similarity of two representations in latent entity space, where
\begin{equation*}
\SigmoidFn{t} = \frac{1}{1 + \ExpFn{-t}}
\end{equation*}
denotes the sigmoid function and $\SimilarityRandomVariable{}$ is an indicator binary random variable that says whether $\TargetEntity{}$ is similar to $\IndexedNGramProjection{}$.

We then approximate the probability of an entity $\TargetEntity{}$ given an n-gram by randomly sampling $\NumNegativeExamples{}$ contrastive examples:
\newcommand{\UnnormedProbTargetGivenContext}[2][i]{\UnnormedCondProb{\Entity{}_#1}{\NGram{#2}}}
\newcommand{\LogUnnormedProbTargetGivenContext}[2][i]{\log\UnnormedProbTargetGivenContext[#1]{#2}}
\begin{eqnarray}
\label{eq:instance_loss}
\lefteqn{\LogUnnormedProbTargetGivenContext{j,}} \\
& = & \log{\ProbSimilarEntity{\TargetEntityRepr{}}} \nonumber \\
&   & + \sum_{\substack{k=1, \\ \Entity{}_k \sim U(\Entities{})}}^{\NumNegativeExamples{}} \log{\left(1 - \ProbSimilarEntity{\EntityEmbedding{}_k}\right)} \nonumber
\end{eqnarray}
where $U(X)$ denotes the uniform distribution over entities $\Entities{}$, the noise distribution used in NCE \citep{Gutmann2010}. Eq.~\ref{eq:instance_loss} avoids iterating over all entities during parameter estimation as we stochastically sample $\NumNegativeExamples{}$ entities uniformly as negative training examples.\footnote{We exploit the special nature of our evaluation scenario where we know the unique association between documents and entities. The setup can easily be adapted to the more general case where a document is associated with multiple entities by extracting the same word sequences from the document for every associated entity.}

During model construction we maximize the log-probability \eqref{eq:instance_loss} using batched gradient descent. The loss function for a single batch of $\BatchSize{}$ instances $\left((\NGram[]{k,}), \Entity{}_k\right)$ consisting of n-grams sampled from documents $\AssociatedDocuments{\Entity{}_k}$ (see~\S\ref{sec:experimental:design}) and associated entity $\Entity{}_k$ is as follows:
\begin{eqnarray}
\label{eq:loss}
\lefteqn{\LossFn{}}
\nonumber \\
& = & - \frac{1}{\BatchSize{}} \sum_{k = 1}^\BatchSize{} \LogUnnormedProbTargetGivenContext[k]{k,} \nonumber \\
\nonumber \\
&   & + \frac{\lambda}{2 \BatchSize{}} \left( \sum_{i,j} \VocabularyEmbeddingMatrix{}_{i,j}^2 + \sum_{i,j} \EntityEmbeddingMatrix{}_{i,j}^2 + \sum_{i,j} \MappingMatrix{}_{i,j}^2 \right),
\end{eqnarray}
where $\lambda$ is a weight regularization parameter. Instances are shuffled before batches are created. The update rule for a particular parameter $\theta$ ($\VocabularyEmbeddingMatrix{}$, $\EntityEmbeddingMatrix{}$, $\MappingMatrix{}$ or $\MappingBias{}$) given a single batch of size $\BatchSize{}$ is:
\begin{equation}
\label{eq:update}
\theta^{(t+1)} = \theta^{(t)} - \LearningRate{}^{(t)} \odot \frac{\partial \LossFnIdentifier}{\partial \theta}\LossFnArguments[^{(t)}]{},
\end{equation}
where $\LearningRate{}^{(t)}$ and $\theta^{(t)}$ denote the per-parameter learning rate and parameter $\theta$ at time $t$, respectively. The learning rate $\LearningRate{}$ consists of the same number of elements as there are parameters; in the case of a global learning rate, all elements of $\LearningRate{}$ are equal to each other.
The derivatives of the loss function \eqref{eq:loss} are given in the Appendix.

%% file: experiments.tex

\section{Experimental Setup}
\label{section:setup}

\subsection{Research questions}
\label{section:setup:rq}

In this paper we investigate the problem of constructing a latent vector model of words and entities by directly modeling the discriminative relation between entities and word context. We seek to answer the following research questions:

\newcommand{\RQ}[2]{
	\begin{description}[topsep=0pt]
	\phantomsection\label{section:setup:rq#1}
	\item[RQ#1] #2
	\end{description}
}

\newcommand{\RQRef}[1]{\textbf{\hyperref[section:setup:rq#1]{RQ#1}}}

\RQ{1}{\ResearchQuestionOne{}}
In \S\ref{section:methodology} we introduced various hyper-parameters along with the definition of \FullModelName{}. We have the size of word representations $\WordEmbeddingSize{}$ and the dimensionality of the entity representations $\EntityEmbeddingSize{}$. During parameter estimation, the window size $\WindowSize{}$ influences the context width presented as evidence for a particular entity. What is the influence of these parameters on the effectiveness of \ModelName{} and can we identify relations among parameters?

\RQ{2}{\ResearchQuestionTwo{}}
Is there a single method that always performs best or does effectiveness differ per domain? Does an increase in the vector space dimensionality impact the effectiveness of these methods?
\RQ{3}{\ResearchQuestionThree{}}
How does \ModelName{} compare to language models on a per-topic basis? Are there particular topics that work especially well with either type of ranker?
\RQ{4}{\ResearchQuestionFour{}}
What if we combine popularity-based, exact matching and latent vector space features in a linear learning-to-rank setting? Do we observe an increase in effectiveness if we combine these features?

\subsection{Experimental design}
\label{sec:experimental:design}

\newcommand{\NoEmphHomeKitchen}{Home \& Kitchen}
\newcommand{\NoEmphClothing}{Clothing, Shoes \& Jewelry}
\newcommand{\NoEmphPetSupplies}{Pet Supplies}
\newcommand{\NoEmphSports}{Sports \& Outdoors}

\newcommand{\HomeKitchen}{\emph{\NoEmphHomeKitchen{}}}
\newcommand{\Clothing}{\emph{\NoEmphClothing{}}}
\newcommand{\PetSupplies}{\emph{\NoEmphPetSupplies{}}}
\newcommand{\Sports}{\emph{\NoEmphSports{}}}

To answer the research questions posed in \S\ref{section:setup:rq}, we evaluate \ModelName{} in an entity retrieval setting organized around Amazon products (see \S\ref{sec:benchmarks}). We choose to experiment with samples of Amazon product data \citep{McAuley2015-1,McAuley2015-2} for the following reasons:
\begin{inparaenum}[(1)]
	\item The collection contains heterogeneous types of evidential documents associated with every entity: descriptions as well as reviews.
	\item Every department (e.g., \HomeKitchen{}) constitutes a separate, self-contained domain.
	\item Within each department there is a hierarchical taxonomy that partitions the space of entities in a rich structure. We can use the labels associated with these partitions and the partitions themselves as ground truth during evaluation.
	\item Every department consists of a large number of products categorized over a large number of categories. Importantly, this allows us to construct benchmarks with an increasing number of entities.
	\item Every product has a variety of attributes that can be used as popularity-based features in a learning-to-rank setting.
\end{inparaenum}

\newcommand{\EntityFindingBenchmarksNoRef}{\HomeKitchen{}, \Clothing{}, \PetSupplies{} and \Sports{} product search benchmarks}
\newcommand{\EntityFindingBenchmarks}{\EntityFindingBenchmarksNoRef{} (\S\ref{sec:benchmarks})}

\begin{table*}[th]
\centering

\caption{Overview of the \EntityFindingBenchmarksNoRef{}. T and V denote the test and validation sets, respectively. Arithmetic mean and standard deviation are reported wherever applicable.\label{tbl:benchmarks}}

\IfFileExists{resources/statistics.tex}{
\input{resources/statistics.tex}}{
\resizebox{\textwidth}{!}{\missingfigure{Statistics}}}

\end{table*}

To answer \RQRef{1} we investigate the relation between the dimensionality of the entity representations $\EntityEmbeddingSize{}$ and window size $\WindowSize{}$. The latter, the window size $\WindowSize{}$, controls the context width the model can learn from, while the former, the dimensionality of the entity representations $\EntityEmbeddingSize{}$, influences the number of parameters and expressive power of the model. We sweep exponentially over $\WindowSize{}$ ($2^i$ for $0 \leq i < 6$) and $\EntityEmbeddingSize{}$ ($2^i$ for $6 \leq i < 11$). \RQRef{2} is answered by comparing \ModelName{} with latent vector space model baselines (\S\ref{section:baselines}) for an increasing entity space dimensionality $\EntityEmbeddingSize{}$ ($2^i$ for $6 \leq i < 11$). For \RQRef{3}, we compare the per-topic paired differences between \ModelName{} and a lexical language model. In addition, we investigate the correlation between lexical matches in relevant entity documents and ranker preference. We address \RQRef{4} by evaluating \ModelName{} as a feature in a machine-learned ranking in addition to query-independent and lexical features.

\newcommand{\m}{\sqrt{\frac{6.0}{m + n}}}

The number of $\WindowSize{}$-grams sampled per entity $\Entity{} \in \Entities{}$ from associated documents $\AssociatedDocuments{\Entity{}}$ in every epoch (i.e., iteration of the training data) is equal to
$%
\ceil*{\frac{1}{\Length{\Entities{}}} \sum_{\Document{} \in \Documents{}} \text{max}\left(\Length{\Document{}} - \WindowSize{} + 1, 0\right)},
$%
where the $\Length{\cdot}$ operator is used interchangeably for the size of set $\Entities{}$ and the number of tokens in documents $\Document{} \in \Documents{}$. This implicitly imposes a uniform prior over entities (i.e., stratified sampling where every entity is of equal importance).
The word vocabulary $V$ is created for each benchmark by ignoring punctuation, stop words and case; numbers are replaced by a numerical placeholder token. We prune $V$ by only retaining the $2^{16}$ most-frequent words so that each word can be encoded by a 16-bit unsigned integer.
In terms of parameter initialization of the \FullModelName{} model, we sample the initial matrices $\VocabularyEmbeddingMatrix{}$, $\MappingMatrix{}$ (Eq.~\ref{eq:hash_fn}) and $\EntityEmbeddingMatrix{}$ uniformly in the range
$
\left[ -\m, \m \; \right]
$
for an $m \times n$ matrix, as this initialization scheme is known to improve model training convergence \citep{Glorot2010}, and take the bias vector $\MappingBias{}$ to be null. The number of word features is set to $\WordEmbeddingSize{}=300$, similar to \citep{Mikolov2013-2}.
We take the number of negative examples $\NumNegativeExamples{} = 10$ to be fixed. \citet{Mikolov2013-2} note that a value of $\NumNegativeExamples{}$ between 10 and 20 is sufficient for large data sets \citep{Mnih2013}.

We used Adam ($\alpha = 0.001, \beta_1 = 0.9, \beta_2 = 0.999$) \citep{Kingma2014} with batched gradient descent ($\BatchSize{}=4096$) and weight decay $\lambda=0.01$ during training on NVidia Titan X GPUs. Adam has been designed specifically for non-stationary, stochastic cost functions like the one we defined in Eq.~\ref{eq:instance_loss}. For every model, we iterate over the training data 15 times and choose the best epoch based on the validation sets (Table~\ref{tbl:benchmarks}).

\vfill
\subsection{Product search benchmarks}
\label{sec:benchmarks}

We evaluate on four samples from different product domains\footnote{A list of product identifiers, topics and relevance assessments can be found at \paperImplementationUrl{}.} (Amazon departments), each with of an increasing number of products: \HomeKitchen{} (\numprint{8192} products), \Clothing{} (\numprint{16384} products), \PetSupplies{} (\numprint{32768} products) and \Sports{} (\numprint{65536} products); see Table~\ref{tbl:benchmarks}. The documents associated with every product consist of the product description plus reviews provided by Amazon customers.

\citet[p.~24]{Rowley2000} describes directed product search as users searching for ``a producer's name, a brand or a set of terms which describe the category of the product.'' Following this observation, the test topics $\CategoricalTopic{}_i$ are extracted from the categories each product belongs to. Category hierarchies of less than two levels are ignored, as the first level in the category hierarchy is often non-descriptive for the product (e.g., in \Clothing{} this is the gender for which the clothes are designated). Products belonging to a particular category hierarchy are considered as relevant for its extracted topic. Products can be relevant for multiple topics. Textual representations $\Query{}_{\CategoricalTopic{}_i}$ of the topics based on the categories are extracted as follows. For a single hierarchy of categories, we tokenize the titles of its sub-categories and remove stopwords and duplicate words. For example, a digital camera lense found in the \emph{Electronics} department under the categorical topic \emph{Camera \& Photo} $\rightarrow$ \emph{Digital Camera Lenses} will be relevant for the textual query ``\emph{photo camera lenses digital}.'' Thus, we only have two levels of relevance. We do not index the categories of the products as otherwise the query would match the category and retrieval would be trivial.

\subsection{Evaluation measures and significance}
\label{section:evaluation}
\newcommand{\DCG}{DCG}
\newcommand{\IdealDCG}{perfect \DCG{}}
\newcommand{\IdealDCGCut}{\IdealDCG{}@100}
\newcommand{\NDCG}{N\DCG{}}
\newcommand{\NDCGCut}{\NDCG{}@100}
\newcommand{\MAPCut}{MAP@100}
\newcommand{\PrecisionCut}{Precision@k}

\newcommand{\Significant}{^{*}}
\newcommand{\MoreSignificant}{^{**}}
\newcommand{\HighlySignificant}{^{***}}

To measure retrieval effectiveness, we report Normalized Discounted Cumulative Gain (\NDCG{}). For \RQRef{4}, we additionally report \PrecisionCut{} ($k = 5, 10$).
Unless mentioned otherwise, significance of observed differences is determined using a two-tailed paired Student's t-test \citep{Smucker2007} ($\HighlySignificant{} \, p < 0.01$;  $\MoreSignificant{} \, p < 0.05$; $\Significant{} \, p < 0.1$).

\subsection{Methods used in comparisons}
\label{sec:other_methods}

We compare \FullModelName{} to state-of-the-art latent vector space models for entity retrieval that are known to perform semantic matching \citep{Li2014}. We also conduct a contrastive analysis between \ModelName{} and smoothed language models with exact matching capabilities.
\paragraph*{Vector Space Models for entity finding}
\label{section:baselines}
\newcommand{\RelevantDocuments}[1]{\text{rel}_{#1}}
\newcommand{\Ideal}[1]{#1^*}
\citet{Demartini2009} propose a formal model for finding entities using document vector space models (\S\ref{section:methodology:background}). We compare the retrieval effectiveness of \ModelName{} with baseline latent vector space models created using
\begin{inparaenum}[(1)]
	\item Latent Semantic Indexing (\LSI{}) \citep{Deerwester1990} with TF-IDF term weighting,
	\item Latent Dirichlet Allocation (\LDA{}) \citep{Blei2013} with $\alpha = \beta = 0.1$, where a document is represented by its topic distribution, and
	\item \WordToVec{} \citep{Mikolov2013} with CBOW and negative sampling, where a query/document is represented by the average of its word embeddings (same for queries in \ModelName{}). Similar to \ModelName{}, we train \WordToVec{} for 15 iterations and select the best-performing model using the validation sets (Table~\ref{tbl:benchmarks}).
\end{inparaenum}
\paragraph*{\QueryLikelihoodLM{}}
\label{section:baselines:lm}
For every entity a pro\-file-based statistical language model is constructed using maximum-likelihood estimation \citep{Vapnik1998,Liu2005,Balog2006}, which is then smoothed by the language model of the entire corpus. The retrieval score of entity $\Entity{}$ for query $\Query{}$ is defined as
\begin{equation}
\UnnormedCondProb{\Query{}}{\Entity{}} = \prod_{\Term{}_i \in \Query{}} P(\Term{}_i \mid \theta_{\Entity{}}),
\end{equation}
where $\CondProb{\Term{}}{\theta_{\Entity{}}}$ is the probability of term $\Term{}$ occurring in the smooth\-ed language model of $\Entity{}$ (Jelinek-Mercer smoothing \citep{Zhai2004}). Given a query $\Query{}$, entities are ranked according to $\UnnormedCondProb{\Query{}}{\Entity{}}$ in descending order.

\paragraph*{Machine-learned ranking}
\label{sec:experimental:ltr}
RankSVM models~\citep{Joachims2002} in \S\ref{section:discussion:ltr} and~\ref{section:analysis:representations} are trained using stochastic gradient descent using the implementation of \citet{Sculley2009}. We use default values for all parameters, unless stated otherwise. For the experiment investigating \ModelName{} as a feature in machine-learn\-ed ranking in \S\ref{section:discussion:ltr}, we construct training examples by using the relevant entities as positive examples. Negative instances are generated by sampling from the non-relevant entities with replacement until the class distribution is uniform.

%% file: resources/statistics.tex
\begin{tabular}{lcccc}

\toprule
& \NoEmphHomeKitchen{} & \NoEmphClothing{} & \NoEmphPetSupplies{} & \NoEmphSports{} \\

\midrule

\textbf{Corpus} (train) \\

Number of documents & \numprint{88130}  & \numprint{94024}  & \numprint{416993}  & \numprint{502313} \\
Document length & \numprint{70.02} $\pm\, 73.82$  & \numprint{58.41} $\pm\, 61.90$  & \numprint{77.48} $\pm\, 78.44$  & \numprint{72.52} $\pm\, 81.47$ \\

\\

Number of entities & \numprint{8192}  & \numprint{16384}  & \numprint{32768}  & \numprint{65536} \\
Documents per entity & \numprint{10.76} $\pm\, 52.01$  & \numprint{5.74} $\pm\, 18.60$  & \numprint{12.73} $\pm\, 55.98$  & \numprint{7.66} $\pm\, 30.38$ \\

\midrule

\textbf{Topics} (test) \\

Topics & \MultiRowCell{\numprint{657} (T)\\ \numprint{72} (V)}  & \MultiRowCell{\numprint{750} (T)\\ \numprint{83} (V)}  & \MultiRowCell{\numprint{385} (T)\\ \numprint{42} (V)}  & \MultiRowCell{\numprint{1879} (T)\\ \numprint{208} (V)} \\
\\

Terms per topic & \numprint{5.11} $\pm\, 1.79$  & \numprint{4.10} $\pm\, 1.86$  & \numprint{3.73} $\pm\, 1.62$  & \numprint{4.64} $\pm\, 1.68$ \\

\\

\MultiRowCell{Relevant entities\\per topic} & \MultiRowCell{\numprint{10.92} $\pm\, 32.41$ (T)\\ \numprint{10.29} $\pm\, 15.66$ (V)} & \MultiRowCell{\numprint{20.15} $\pm\, 57.78$ (T)\\ \numprint{12.13} $\pm\, 19.85$ (V)} & \MultiRowCell{\numprint{75.96} $\pm\, 194.44$ (T)\\ \numprint{57.40} $\pm\, 88.91$ (V)} & \MultiRowCell{\numprint{29.27} $\pm\, 61.71$ (T)\\ \numprint{38.25} $\pm\, 157.34$ (V)}\\

\bottomrule

\end{tabular}

%% file: discussion.tex

\newcommand{\RQAnswer}[2]{
	\RQRef{#1}: #2%
}

\section{Results and discussion}
\label{section:discussion}

We start by giving a high-level overview of our experimental results (\RQRef{1} and \RQRef{2}), followed by a comparison with lexical matching methods (\RQRef{3}) and the use of \ModelName{} as a ranking feature (\RQRef{4}) (see \S\ref{sec:experimental:design} for an overview of the experimental design).

\subsection{Overview of experimental results}

\begin{figure*}[t]

\newcommand{\heatmap}[4]{%
	\def \inner {%
		\def \PlotPath {resources/heat_map/#1_#2_#3.pdf}%
		\IfFileExists{\PlotPath}{
			\includegraphics[width=0.225\textwidth]{\PlotPath}}{
			\resizebox{0.225\textwidth}{!}{\missingfigure{#3}}}%
	}%
	\ifstrequal{#4}{}{%
	\subfloat{\inner}%
	}{\subfloat[#4\label{fig:heatmap:#2:#1}]{\inner}}%
}

\heatmap{home_and_kitchen}{qrel_validation}{ndcg}{\NoEmphHomeKitchen{}}
\hfill
\heatmap{clothing_shoes_and_jewelry}{qrel_validation}{ndcg}{\NoEmphClothing{}}
\hfill
\heatmap{pet_supplies}{qrel_validation}{ndcg}{\NoEmphPetSupplies{}}
\hfill
\heatmap{sports_and_outdoors}{qrel_validation}{ndcg}{\NoEmphSports{}}

\medskip
\caption{Sensitivity analysis of \ModelName{} in terms of \NDCG{} for window size $\WindowSize{}$ and the size of entity representations $\EntityEmbeddingSize{}$ during parameter estimation (Eq.~\ref{eq:loss}) for models trained on \EntityFindingBenchmarks{} and evaluated on the \textbf{validation} sets.\label{fig:heatmap}}

\vspace*{-.5\baselineskip}
\end{figure*}

\RQAnswer{1}{Fig.~\ref{fig:heatmap} depicts a heat map for every combination of window size and entity space dimensionality evaluated on the validation sets (Table~\ref{tbl:benchmarks}). Fig.~\ref{fig:heatmap} shows that neither extreme values for the dimensionality of the entity representations nor the context width alone achieve the highest performance on the validation sets.

Instead, a low-dimensional entity space (\numprint{128}- and \numprint{256}-dimen\-sional) combined with a medium-sized context window (\numprint{4}- and \numprint{8}-grams) achieve the highest NDCG. In the two largest benchmarks (Fig.~\ref{fig:heatmap:qrel_validation:pet_supplies},~\ref{fig:heatmap:qrel_validation:sports_and_outdoors}) we see that for \numprint{16}-grams, NDCG actually lowers as the dimensionality of the entity space increases. This is due to the model \emph{fitting} the optimization objective (Eq.~\ref{eq:loss}), which we use as an unsupervised surrogate of relevance, too well. That is, as the model is given more learning capacity (i.e., higher dimensional representations), it starts to learn more regularities of natural language which counteract retrieval performance.}

\begin{figure*}[t]

\newcommand{\vectorspace}[4]{
	\subfloat[#4\label{fig:vectorspace:#2:#1}]{%
		\def \PlotPath {resources/vector_space/#1_#2_#3.pdf}%
		\IfFileExists{\PlotPath}{
			\includegraphics[width=0.230\textwidth]{\PlotPath}}{
			\resizebox{0.230\textwidth}{!}{\missingfigure{#4}}}
	}
}

\vectorspace{home_and_kitchen}{qrel_test}{ndcg}{\NoEmphHomeKitchen{}}
\hfill
\vectorspace{clothing_shoes_and_jewelry}{qrel_test}{ndcg}{\NoEmphClothing{}}
\hfill
\vectorspace{pet_supplies}{qrel_test}{ndcg}{\NoEmphPetSupplies{}}
\hfill
\vectorspace{sports_and_outdoors}{qrel_test}{ndcg}{\NoEmphSports{}}

\medskip
\caption{Comparison of \ModelName{} (with window size $\WindowSize{} = 4$) with latent vector space baselines (\LSI{}, \LDA{} and \WordToVec{}; \S\ref{section:baselines}) on \EntityFindingBenchmarks{} and evaluated on the \textbf{test} sets.
Significance (\S\ref{section:evaluation}) is computed between \ModelName{} and the baselines for each vector space size.\label{fig:vectorspace}}

\vspace*{-.5\baselineskip}
\end{figure*}

\RQAnswer{2}{Fig.~\ref{fig:vectorspace} presents a comparison between \ModelName{} (window size $\WindowSize{} = 4$) and vector space model baselines (\S\ref{section:baselines}) for increasing entity representation dimensionality ($2^i$ for $6 \leq i < 11$) on the test sets. \ModelName{} significantly outperforms ($p < 0.01$) all baseline methods in most cases (except for Fig.~\ref{fig:vectorspace:qrel_test:home_and_kitchen} where $\EntityEmbeddingSize = 1024$). For the smaller benchmarks (Fig.~\ref{fig:vectorspace:qrel_test:home_and_kitchen},~\ref{fig:vectorspace:qrel_test:clothing_shoes_and_jewelry}), we see \LSI{} as the main competitor of \ModelName{}. However, as the training corpora become larger (in Fig.~\ref{fig:vectorspace:qrel_test:pet_supplies},~\ref{fig:vectorspace:qrel_test:sports_and_outdoors}), \WordToVec{} outperforms \LSI{} and becomes the main contester of \ModelName{}. On all benchmarks, \ModelName{} peaks when the entity representations are low-dimensional (\numprint{128}- or \numprint{256}-dimensional) and afterwards (for a higher dimensionality) performance decreases. On the other hand, \WordToVec{} stagnates in terms of \NDCG{} around representations of \numprint{512} dimensions and never achieves the same level as \ModelName{} did for one or two orders of magnitude (base 2) smaller representations. This is a beneficial trait of \ModelName{}, as high-dimensional vector spaces are undesirable due to their high computational cost during retrieval \citep{Weber1998}.}

\subsection{A feature for machine-learned ranking}
\label{section:discussion:ltr}

We now investigate the use of \ModelName{} as a feature in a learning to rank setting \citep{Liu2011}. Latent vector space models are known to provide a means of semantic matching as opposed to a purely lexical matching~\citep{Li2014,VanGysel2016}. To determine to which degree this is indeed the case, we first perform a topic-wise comparison between \ModelName{} and a lexical language model, the \QueryLikelihoodLM{} (\QLM{}) \citep{Zhai2004}, as described in \S\ref{section:baselines:lm}. We optimize the parameters of \ModelName{} and \QLM{} on the validation sets for every benchmark (Table~\ref{tbl:benchmarks}). In the case of \ModelName{}, we select the model that performs best in Fig.~\ref{fig:heatmap}. For \QLM{}, we sweep over $\lambda$ linearly from $0.0$ to $1.0$ (inclusive) with increments of $0.05$.

\begin{figure}[bh!]
\vspace*{2em}

\newcommand{\innerdeltaplot}[5]{
	\def \PlotPath {resources/delta/#1_#2_#3_#4_#5_differences.pdf}%
		\IfFileExists{\PlotPath}{
			\includegraphics[width=0.225\textwidth]{\PlotPath}}{
			\resizebox{0.225\textwidth}{!}{\missingfigure{#5}}}
}

\newcommand{\deltaplot}[5]{
	\subfloat[#5\label{fig:delta:#2:#1}]{%
		\innerdeltaplot{#1}{#2}{#3}{#4}{ndcg}%
	}
}

\def \ModelConfiguration{4window.deep_pre_scalable_inproduct_tanh_512embedding_instance}
\newcommand{\BaselineModel}[1]{baseline_model_mle_lm.mle_lm_jm_#1_word_lm_uniform_prior}

\deltaplot{home_and_kitchen}{4window.deep_pre_scalable_inproduct_tanh_128embedding_instance}{\BaselineModel{0.15}}{qrel_test}{\NoEmphHomeKitchen{}}%
\hfill%
\deltaplot{clothing_shoes_and_jewelry}{4window.deep_pre_scalable_inproduct_tanh_1024embedding_instance}{\BaselineModel{0.15}}{qrel_test}{\NoEmphClothing{}}%
\hfill%
\deltaplot{pet_supplies}{4window.deep_pre_scalable_inproduct_tanh_256embedding_instance}{\BaselineModel{0.4}}{qrel_test}{\NoEmphPetSupplies{}}%
\hfill%
\deltaplot{sports_and_outdoors}{8window.deep_pre_scalable_inproduct_tanh_128embedding_instance}{\BaselineModel{0.1}}{qrel_test}{\NoEmphSports{}}%

\medskip
\caption{Per-topic paired differences between \ModelName{} and \QueryLikelihoodLM{} for models trained on \EntityFindingBenchmarks{} and evaluated on the \textbf{test} sets. For every plot, the y-axis indicates $\Delta \text{\NDCG{}}$ between \ModelName{} and a \QueryLikelihoodLM{}. The x-axis lists the topics in the referenced benchmark in decreasing order of $\Delta \text{\NDCG{}}$ such that topics for which \ModelName{} performs better are on the left and vice versa for the \QueryLikelihoodLM{} on the right.\label{fig:delta}}

\vfill
\end{figure}

\begin{table}
	\centering

	\caption{Correlation coefficients between average IDF of lexically matched terms in documents associated with relevant entities and $\bigtriangleup{} \text{\NDCG{}}$. A negative correlation coefficient implies that queries consisting of more specific terms (i.e., low document freq.) that occur exactly in documents associated with relevant entities are more likely to benefit from \QLM{}, whereas other queries (with less specific terms or less exact matches) gain more from \ModelName{}. Significance is achieved for all benchmarks ($p < 0.01$) using a permutation test.\label{tbl:correlations}}

	\begin{tabular}{l c c}

	\toprule
	\textbf{Benchmark} & \textbf{Spearman $R$} & \textbf{Pearson $R$} \\
	\midrule

	\NoEmphHomeKitchen{} & \nprounddigits{2} \npdecimalsign{.} \numprint{-0.301271661356} & \nprounddigits{2} \npdecimalsign{.} \numprint{-0.348547810194} \\
	\NoEmphClothing{} & \nprounddigits{2} \npdecimalsign{.} \numprint{-0.396222193487} & \nprounddigits{2} \npdecimalsign{.} \numprint{-0.369019393753} \\
	\NoEmphPetSupplies{} & \nprounddigits{2} \npdecimalsign{.} \numprint{-0.165627764962} & \nprounddigits{2} \npdecimalsign{.} \numprint{-0.171408122876} \\
	\NoEmphSports{} & \nprounddigits{2} \npdecimalsign{.} \numprint{-0.338266676522} & \nprounddigits{2} \npdecimalsign{.} \numprint{-0.361384861078} \\

	\midrule

	\end{tabular}
\end{table}

\RQAnswer{3}{Fig.~\ref{fig:delta} shows the per-topic paired difference between \ModelName{} and \QLM{} in terms of \NDCG{}. Topics that benefit more from \ModelName{} have a positive value on the y-axis, while those that prefer \QLM{} have a negative value. We can see that both methods perform similarly for many topics (where $\bigtriangleup = 0.0$). For certain topics one method performs substantially better than the other, suggesting that the two are complementary. To further quantify this, we investigate the relation between specific topic terms and their occurrence in documents relevant to these topics. That is, we measure the correlation between the per-topic $\bigtriangleup \text{\NDCG{}}$ (as described above) and the average inverse document frequency (IDF) of exact/lexically matched terms in the profile-based language model. In Table~\ref{tbl:correlations} we observe that queries that contain specific tokens (i.e., with high inverse document frequency) and occur exactly in documents associated with relevant products, benefit more from \QLM{} (lexical matches). Conversely, queries with less specific terms or without exact matches in the profiles of relevant products gain more from \ModelName{} (semantic matches).

This observation motivates the use of \ModelName{} as a ranking feature in addition to traditional language models.}
Specifically, we now evaluate the use of \ModelName{} as a feature in a linear \RankSVM{} (\S\ref{sec:experimental:ltr}). Following \citet{Fang2010}, we consider query-independent (QI) popularity-based features in addition to features provided by \ModelName{} and \QLM{}. This allows us to consider the effect of the query-dependent features independent from their ability to model a popularity prior over entities. Table~\ref{fig:ltr-overview} lists the feature sets.

\begin{table}[t]
	\centering

	\caption{Overview of the feature sets used in the machine-learned ranking experiments.\label{fig:ltr-overview}}

	\small
	\begin{tabular}{c L{6.5cm}}

	\toprule
	\textbf{Features} & \textbf{Description} \\
	\midrule

	\textbf{QI} &
		Query-independent features:%
		\begin{inparaenum}[(1)]
			\item product price;
			\item product description length;
			\item reciprocal of the Amazon sales rank; and
			\item product PageRank scores based on four related product graphs (also bought, also viewed, bought together, buy after viewing).
		\end{inparaenum}
		\\
		\midrule
	\textbf{QLM} & \QueryLikelihoodLM{} using Jelinek-Mercer smoothing with $\lambda$ optimized on the validation set (Table~\ref{tbl:benchmarks}). Posterior $\CondProb{\Query{}}{\Entity{}}$ is used as a feature for entity $\Entity{}$ and query $\Query{}$. \\
	\midrule
	\textbf{\ModelName{}} & \FullModelName{} optimized on the validation set (Table~\ref{tbl:benchmarks}, Fig.~\ref{fig:heatmap}). Similarity $\SimilarityFn{}(\ProjectedQuery{}, \EntityEmbedding{})$ is used as a feature for entity $\Entity{}$, with vector representation $\EntityEmbedding{}$, and query $\Query{}$. \\
	\bottomrule
	\end{tabular}

\end{table}

\begin{table}[t]
	\centering
	\caption{Ranking performance results for query independent (QI) features, the \QueryLikelihoodLM{} (QLM) match feature, the \FullModelName{} (\ModelName{}) match feature and combinations thereof, weighted using \RankSVM{} (\S\ref{section:discussion:ltr}), evaluated on the test sets using 10-fold cross validation, for \EntityFindingBenchmarks{}. The hyperparameters of the individual query features (QLM and LSE) were optimized using the validation sets.
	Significance of the results (\S\ref{section:evaluation}) is computed between {QI + QLM + LSE} and {QI + QLM}.\label{tbl:ltr}}
	\IfFileExists{resources/results-ltr.tex}{
	\small
	\input{resources/results-ltr.tex}}{
	\resizebox{\textwidth}{!}{\missingfigure{LTR}}}
	\vspace*{-.5\baselineskip}
\end{table}

\RQAnswer{4}{Table~\ref{tbl:ltr} shows the results for different combinations of feature sets used in a machine-learned ranker, RankSVM. The experiment was performed using 10-fold cross validation on the test sets (Table~\ref{tbl:benchmarks}). The combination using all features outperforms smaller subsets of features, on all metrics. We conclude that \FullModelName{} adds a signal that is complementary to traditional (lexical) language models, which makes it applicable in a wide range of entity-oriented search engines that use ranker fusion techniques.}

\vfill
\pagebreak
\section{Analysis of representations}
\label{section:analysis:representations}

\newcommand{\IdealEntityEmbedding}{\EntityEmbedding{}^{*}_\CategoricalTopic{}}
\newcommand{\Approximate}[1]{\tilde{#1}}
\newcommand{\ApproximateIdealEntityEmbedding}{\tilde{\EntityEmbedding{}}^{*}_\CategoricalTopic{}}

Next, we analyze the entity representations $\EntityEmbedding{}_i$ of the vector space models independent of the textual representations by providing empirical lower-bounds on their maximal retrieval performance, followed by a comparison with their actual performance so as to measure the effectiveness of word-to-entity mapping $\HashFn{}$.

Fig.~\ref{fig:heatmap}~and~\ref{fig:vectorspace} show which levels of performance may be achieved by using the latent models to generate a ranking from textual queries (Eq.~\ref{eq:hash_fn}). But this is only one perspective. As entities are ranked according to their similarity with the projected query vector $\ProjectedQuery[_{\CategoricalTopic{}}]{}{}$, the performance for retrieving entities w.r.t.\ the textual representation of a topic $\CategoricalTopic{}$ depends on the structure of the entity space $\EntitySpace{}$, the ideal retrieval vector $\IdealEntityEmbedding{} \in \EntitySpace{}$ (i.e., the vector that optimizes retrieval performance), and the similarity between $\ProjectedQuery[_{\CategoricalTopic{}}]{}$ and $\IdealEntityEmbedding{}$.

How can we determine the ideal vector $\IdealEntityEmbedding{}$? First, we define it to be the vector for which the cosine similarity with each of the entity embeddings results in a ranking where relevant entities are ranked higher than non-relevant or unjudged entities. We approximate $\IdealEntityEmbedding{}$ by optimizing the pair-wise SVM objective \citep{Joachims2002,Sculley2009}. That is, for every topic $\CategoricalTopic{}$ we construct a separate RankSVM model based on its ground-truth as follows. We only consider topics with at least two relevant entities, as topics with a single relevant entity have a trivial optimal retrieval vector (the entity representation of the single relevant entity). Using the notation of \citep{Joachims2002}, the normalized entity representations are used as features, and hence the feature mapping $\phi$ is defined as
\begin{equation*}
\phi(\CategoricalTopic{}, \Entity{}_i) = \frac{\EntityEmbedding{}_i}{\norm{\EntityEmbedding{}_i}_2} \text{ for all }\Entity{}_i \in \Entities{}.
\end{equation*}
The target ranking $r^*_\CategoricalTopic{}$ is given by the entities relevant to topic $\CategoricalTopic{}$. Thus, the features for every entity become the entity's normalized representation and its label is positive if it is relevant for the topic and negative otherwise. The pair-wise objective then finds a weight vector such that the ranking generated by ordering according to the vector scalar product between the weight vector and the normalized entity representations correlates with the target ranking $r^*_\CategoricalTopic{}$. Thus, our approximation of the \emph{ideal vector}, $\ApproximateIdealEntityEmbedding$, is given by the weight vector $w_\CategoricalTopic{}$ for every $\CategoricalTopic{}$.\footnote{Note that $\ApproximateIdealEntityEmbedding{}$ does not take into account the textual representations $\Query{}_\CategoricalTopic{}$ of topic $\CategoricalTopic{}$, but only the clustering of entities relevant to $\CategoricalTopic{}$ and their relation to other entities.}

\newcommand{\FullVersion}{false}

\begin{figure\if\FullVersion{}*\fi}[t]

\newcommand{\centroids}[4]{%
	\def \inner {%
		\def \PlotPath {resources/centroids/#1_#2_#3.pdf}%
		\IfFileExists{\PlotPath}{
			\includegraphics[width=0.225\textwidth]{\PlotPath}}{
			\resizebox{0.225\textwidth}{!}{\missingfigure{#4}}}
	}%
	\ifstrequal{#4}{}{%
	\subfloat{\inner}%
	}{\subfloat[#4\label{fig:heatmap:#2:#1}]{\inner}}%
}

\centroids{home_and_kitchen}{absolute}{ndcg}{\NoEmphHomeKitchen{}}
\if\FullVersion{}
\hfill
\centroids{clothing_shoes_and_jewelry}{absolute}{ndcg}{}
\fi
\hfill
\centroids{pet_supplies}{absolute}{ndcg}{\NoEmphPetSupplies{}}
\if\FullVersion{}
\hfill
\centroids{sports_and_outdoors}{absolute}{ndcg}{}
\fi

\medskip
\caption{Comparison of the approximately ideal retrieval vector $\ApproximateIdealEntityEmbedding{}$ with the projected query retrieval vector $\ProjectedQuery{}$ for latent entity models built using \ModelName{}, \LSI{}, \LDA{} and \WordToVec{} (\S\ref{section:baselines}) on \if\FullVersion{}\EntityFindingBenchmarks{}\else{}\HomeKitchen{} and \PetSupplies{} product search benchmarks (\S\ref{sec:benchmarks})\fi{} and evaluated on the \textbf{test} sets\if\FullVersion{}\else{}. The plots for \Clothing{} and \Sports{} product search benchmarks are qualitatively similar to the ones shown\fi{}.
The figures show the absolute performance in terms of \NDCG{} of $\ApproximateIdealEntityEmbedding{}$ (dashed curves) and $\ProjectedQuery{}$ (solid curves); significance (\S\ref{section:evaluation}) for the results for the approximately ideal retrieval vectors $\ApproximateIdealEntityEmbedding{}$ is computed between \ModelName{} and the best-performing baseline for each vector space size and indicated along the x-axis.
\label{fig:idealcentroids}}

\end{figure\if\FullVersion{}*\fi}

What is the performance of this approximately ideal vector representation? And how far are our representations removed from it? Fig.~\ref{fig:idealcentroids} shows the absolute performance of $\ApproximateIdealEntityEmbedding{}$ (dashed curves) and $\ProjectedQuery{}$ (solid curves) in terms of \NDCG{}. Comparing the (absolute) difference between every pair of dashed and solid curves for a single latent model gives an intuition of how much performance in terms of \NDCG{} there is to gain by improving the projection function $\HashFn{}$ for that method. The approximately ideal vectors $\ApproximateIdealEntityEmbedding{}$ discovered for \ModelName{} outperform all baselines significantly. Interestingly, for representations created using \LDA{}, the optimal performance goes up while the actual performance stagnates. This indicates that a higher vector space dimensionality renders better representations using \LDA{}, however, the projection function $\HashFn{}$ is unable to keep up in the sense that projected query vectors are not similar to the representations of their relevant entities. The latent models with the best representations (\ModelName{} and \LSI{}) also have the biggest gap between $\ProjectedQuery{}$ and $\ApproximateIdealEntityEmbedding{}$ in terms of achieved \NDCG{}.

We interpret the outcomes of our analysis as follows. The entity space $\EntitySpace{}$ has more degrees of freedom to cluster entities more appropriately as the dimensionality of $\EntitySpace{}$ increases. Consequently, the query projection function $\HashFn{}$ is expected to learn a more complex function. In addition, as the dimensionality of $\EntitySpace{}$ increases, so does the modeling capacity of the projection function $\HashFn{}$ in the case of \ModelName{} and \LSI{} (i.e., the transformation matrices become larger) and therefore more parameters have to be learned. We conclude that our method can more effectively represent entities in a lower-dimensional space than \LSI{} by making better use of the vector space capacity. This is highly desirable, as the asymptotic runtime complexity of many algorithms operating on vector spaces increases at least linearly \citep{Weber1998} with the size of the vectors.

%% file: resources/results-ltr.tex
\begin{tabular}{c@{ }c@{ }c@{ }c}
\toprule
\multirow{2}{*}{} & \multicolumn{3}{c}{Home \& Kitchen} \\ 
& NDCG & P@5 & P@10 \\ 
\cmidrule(lr){2-4}
\multicolumn{1}{l}{QI} & $\phantom{}0.005$ & $\phantom{}0.002$ & $\phantom{}0.001$ \\ 
\multicolumn{1}{l}{QI + QLM} & $\phantom{}0.321$ & $\phantom{}0.180$ & $\phantom{}0.145$ \\ 
\multicolumn{1}{l}{QI + LSE} & $\phantom{}0.257$ & $\phantom{}0.121$ & $\phantom{}0.107$ \\ 
\multicolumn{1}{l}{QI + QLM + LSE} & $\phantom{\HighlySignificant{}}\textbf{0.352}\HighlySignificant{}$ & $\phantom{\MoreSignificant{}}\textbf{0.192}\MoreSignificant{}$ & $\phantom{\HighlySignificant{}}\textbf{0.157}\HighlySignificant{}$ \\ 
\cmidrule{1-4}
\multirow{2}{*}{} & \multicolumn{3}{c}{Clothing, Shoes \& Jewelry} \\ 
& NDCG & P@5 & P@10 \\ 
\cmidrule(lr){2-4}
\multicolumn{1}{l}{QI} & $\phantom{}0.002$ & $\phantom{}0.001$ & $\phantom{}0.001$ \\ 
\multicolumn{1}{l}{QI + QLM} & $\phantom{}0.177$ & $\phantom{}0.079$ & $\phantom{}0.068$ \\ 
\multicolumn{1}{l}{QI + LSE} & $\phantom{}0.144$ & $\phantom{}0.065$ & $\phantom{}0.057$ \\ 
\multicolumn{1}{l}{QI + QLM + LSE} & $\phantom{\HighlySignificant{}}\textbf{0.198}\HighlySignificant{}$ & $\phantom{\HighlySignificant{}}\textbf{0.094}\HighlySignificant{}$ & $\phantom{\HighlySignificant{}}\textbf{0.080}\HighlySignificant{}$ \\ 
\cmidrule{1-4}
\multirow{2}{*}{} & \multicolumn{3}{c}{Pet Supplies} \\ 
& NDCG & P@5 & P@10 \\ 
\cmidrule(lr){2-4}
\multicolumn{1}{l}{QI} & $\phantom{}0.003$ & $\phantom{}0.002$ & $\phantom{}0.002$ \\ 
\multicolumn{1}{l}{QI + QLM} & $\phantom{}0.250$ & $\phantom{}0.212$ & $\phantom{}0.199$ \\ 
\multicolumn{1}{l}{QI + LSE} & $\phantom{}0.268$ & $\phantom{}0.222$ & $\phantom{}0.214$ \\ 
\multicolumn{1}{l}{QI + QLM + LSE} & $\phantom{\HighlySignificant{}}\textbf{0.298}\HighlySignificant{}$ & $\phantom{\HighlySignificant{}}\textbf{0.255}\HighlySignificant{}$ & $\phantom{\HighlySignificant{}}\textbf{0.236}\HighlySignificant{}$ \\ 
\cmidrule{1-4}
\multirow{2}{*}{} & \multicolumn{3}{c}{Sports \& Outdoors} \\ 
& NDCG & P@5 & P@10 \\ 
\cmidrule(lr){2-4}
\multicolumn{1}{l}{QI} & $\phantom{}0.001$ & $\phantom{}0.001$ & $\phantom{}0.001$ \\ 
\multicolumn{1}{l}{QI + QLM} & $\phantom{}0.235$ & $\phantom{}0.183$ & $\phantom{}0.156$ \\ 
\multicolumn{1}{l}{QI + LSE} & $\phantom{}0.188$ & $\phantom{}0.132$ & $\phantom{}0.121$ \\ 
\multicolumn{1}{l}{QI + QLM + LSE} & $\phantom{\HighlySignificant{}}\textbf{0.264}\HighlySignificant{}$ & $\phantom{\HighlySignificant{}}\textbf{0.192}\HighlySignificant{}$ & $\phantom{\HighlySignificant{}}\textbf{0.172}\HighlySignificant{}$ \\ 
\bottomrule
\end{tabular}

%% file: conclusions.tex

\vfill
\pagebreak

\vspace*{-3.5em}
\section{Conclusions}
\label{section:conclusions}

We have introduced \FullModelName{}, an unsupervised latent vector space model for product search. It jointly learns a unidirectional mapping between, and latent vector representations of, words and products. We have also defined a formalism for latent vector space models where latent models are decomposed into a map\-ping from word sequences to the product vector space, representations of products in that space, and a similarity function.
We have evaluated our model using Amazon product data, and compared it to state-of-the-art latent vector space models for product ranking (\LSI{}, \LDA{} and \WordToVec{}). \ModelName{} outperforms all baselines for lower-dimensional vector spaces.

In an analysis of the vector space models, we have compared the performance achieved with the ideal performance of the proposed product representations. We have shown that \ModelName{} constructs better product representations than any of the baselines. In addition, we have obtained important insights w.r.t.\ how much performance there is to gain by improving the individual components of latent vector space models. Future work can focus on improving the mapping from words to products by incorporating specialized features or increasing the mapping's complexity. In addition, semi-supervised learning may help specialize the vector space and mapping function for particular retrieval settings.

A comparison of \ModelName{} with a smoothed lexical language model unveils that the two methods make very different errors. Some directed product search queries require lexical matching, others benefit from the semantic matching capabilities of latent models. We have evaluated \ModelName{} as a feature in a machine-learned ranking setting and found that adding \ModelName{} to language models and popularity-based features significantly improves retrieval performance.

As to future work, in this paper we focus on the unsupervised setting where we have a description and a set of reviews associated with every product. Fig.~\ref{fig:idealcentroids} shows that there is a lot of performance to gain by improving the query projection function $\HashFn{}$. In a semi-supervised setting, the difference between $\IdealEntityEmbedding{}$ and $\ProjectedQuery{}$ can be minimized according to pairs of queries and ideal rankings. As an additional step, query-relevance training data could be incorporated during estimation of the entity space $\EntitySpace{}$. Moreover, as mentioned in \S\ref{section:analysis:representations}, the query projection function $\HashFn{}$ is expected to learn a more complicated mapping. Hence, it may be beneficial to consider incorporating additional hierarchical depth using multiple non-linear transformations in the construction of $\HashFn{}$. More generally, the obtained product representations can be beneficial for various entity-oriented prediction tasks such as entity disambiguation or related entity finding. While we have focused on product retrieval  in this work, the proposed model, insights and ideas can be applied in broader settings, such as entity finding and ad-hoc retrieval.

\smallskip
\begin{spacing}{1}
\noindent\small
\textbf{Acknowledgments.}
The authors would like to thank Artem Grotov, Nikos Voskarides, Zhaochun Ren, Tom Kenter, Manos Tsagkias, Hosein Azar\-bonyad and the anonymous reviewers for their valuable comments and suggestions.
This research was supported by
Ahold,
Amsterdam Data Science,
Blendle,
the Bloomberg Research Grant program,
the Dutch national program COMMIT,
Elsevier,
the European Community's Seventh Framework Programme (FP7/2007-2013) under
grant agreement nr 312827 (VOX-Pol),
the ESF Research Network Program ELIAS,
the Google Faculty Research Award scheme,
the Royal Dutch Academy of Sciences (KNAW) under the Elite Network Shifts project,
the Microsoft Research Ph.D.\ program,
the Netherlands eScience Center under project number 027.012.105,
the Netherlands Institute for Sound and Vision,
the Netherlands Organisation for Scientific Research (NWO)
under pro\-ject nrs
727.\-011.\-005, 
612.001.116, 
HOR-11-10, 
640.006.013, 
612.\-066.\-930, 
CI-14-25, 
SH-322-15, 
652.\-002.\-001, 
612.\-001.\-551, 
652.\-001.\-003, 
and
Yandex.
All content represents the opinion of the authors, which is not necessarily shared or endorsed by their respective employers and/or sponsors.
\end{spacing}

%% file: appendix.tex

\appendix

\let\oldcenter\center
\let\oldendcenter\endcenter
\renewenvironment{center}{\setlength\topsep{0pt}\oldcenter}{\oldendcenter}

\newcommand{\RescaleEqn}[1]{
	\vspace*{-0.0\baselineskip}%
	\begin{center}%
	\scalebox{0.65}{
	\begin{minipage}{\linewidth}
	#1
	\end{minipage}
	}%
	\end{center}%
	\vspace*{-0.0\baselineskip}
}

\newcommand{\LossGradNoReg}[1]{
	\frac{\partial \LossFnIdentifier{}}{
		\partial #1} \LossFnArguments{} = - \frac{1}{m}  \Bigg(
			\sum_{k = 1}^m
			\frac{1}{p_k}
			\frac{\partial p_k}{\partial #1}
		\Bigg)}%

Denote $p_k = \UnnormedProbTargetGivenContext[k]{k,}$. The derivative of \eqref{eq:loss} w.r.t. bias term $\MappingBias{}$ equals

\vspace*{-0.5em}
\RescaleEqn{
\begin{eqnarray*}
\LossGradNoReg{\MappingBias{}}
\end{eqnarray*}
}
\vspace*{-0.5em}

\noindent%
and w.r.t. an arbitrary matrix parameter $\theta$ ($\VocabularyEmbeddingMatrix{}$, $\EntityEmbeddingMatrix{}$ or $\MappingMatrix{}$):

\vspace*{-0.5em}
\RescaleEqn{
\begin{eqnarray*}
\LossGradNoReg{\theta} \nonumber + \frac{\lambda}{m} \sum_{i,j} \theta_{i,j}.
\end{eqnarray*}
\vspace*{-0.5em}
\renewcommand{\ProbSimilarEntity}[1]{\CondProb{\SimilarityRandomVariable{}}{#1, \NGramProjection{}}}%
}

\noindent%
Ignoring the subscripts for batch instances and word positions for ease of notation, for a single instance we denote $\Entity{}^+$ as the target entity and $\Entities{}^-$ as the sample of $\NumNegativeExamples{}$ contrastive negative examples. We have

\vspace*{-0.5em}
\RescaleEqn{
\begin{eqnarray*}
p & = & \ProbSimilarEntity{\EntityEmbedding{}^+} \cdot \prod_{\Entity{}^- \in \Entities{}^-} (1 - \ProbSimilarEntity{\EntityEmbedding{}^-})
\end{eqnarray*}
}
\vspace*{-0.5em}

\noindent%
where application of the product rule in the computation of $\frac{\partial p}{\partial \theta}$ is omitted due to space constraints.

\newcommand{\BigSkip}{\qquad\qquad\qquad\qquad\qquad\qquad\qquad\qquad}

For $\theta$ ($\MappingMatrix{}$, $\MappingBias{}$, $\EntityEmbeddingMatrix{}$ or $\VocabularyEmbeddingMatrix{}$) we observe

\vspace*{-0.5em}
\RescaleEqn{
\begin{eqnarray*}
\lefteqn{\frac{\partial \ProbSimilarEntity{\EntityEmbedding{}}}{\partial \theta} = \ProbSimilarEntity{\EntityEmbedding{}} \cdot} \nonumber \\
&   & (1 - \ProbSimilarEntity{\EntityEmbedding{}}) \cdot \frac{\partial \, \DotProduct{\EntityEmbedding{}}{\NGramProjection{}}}{\partial \theta} \nonumber
\end{eqnarray*}
}
\vspace*{-0.5em}

\noindent%
For a single entity representation $\EntityEmbedding{}$ (a row of matrix $\EntityEmbeddingMatrix{}$),

\vspace*{-0.5em}
\RescaleEqn{
\begin{equation*}
\frac{\partial \, \DotProduct{\EntityEmbedding{}}{\NGramProjection{}}}{\partial \EntityEmbedding{}} = \NGramProjection{}
\end{equation*}
}
\vspace*{-0.5em}

\noindent%
where we observe that the update to an entity representation is the projected representation of the input n-gram multiplied by a scalar.

\newcommand{\dtanh}{\text{sech}^2}

The symbolic derivative of the dot product between the entity representation and the projected n-gram w.r.t.\ bias term $\MappingBias{}$, linear map $\MappingMatrix{}$ and word representations $\VocabularyEmbeddingMatrix{}$, respectively, are:

\vspace*{-1em}
\RescaleEqn{
\begin{flalign*}
&{\frac{\partial \, \DotProduct{\EntityEmbedding{}}{\NGramProjection{}}}{\partial \MappingBias{}} = \EntityEmbedding{} \odot \dtanh{}\!\left(\LinearCombination{}\right)} & \\%
\vspace*{\baselineskip}
&{\frac{\partial \, \DotProduct{\EntityEmbedding{}}{\NGramProjection{}}}{\partial \MappingMatrix{}} = \!\left( \EntityEmbedding{} \odot \dtanh{}\!\left(\LinearCombination{}\right) \right) \cdot \nonumber} \Transpose{\!\left(\AvgWordEmbedding{}\right)} & \\
\vspace*{\baselineskip}
&{\frac{\partial \, \DotProduct{\EntityEmbedding{}}{\NGramProjection{}}}{\partial \VocabularyEmbeddingMatrix{}} = \Transpose{\MappingMatrix{}} \cdot \!\left( \EntityEmbedding{} \odot \dtanh{}\!\left(\LinearCombination{}\right)\right) \cdot \nonumber} \Transpose{\!\left(\NormalizedBoWVector{}\right)} & \\
\end{flalign*}
}